\documentclass[prd,preprint,showpacs,groupedaddress]{revtex4-1}
\usepackage{amsmath}
\usepackage{amsfonts}
\usepackage{amssymb}
\usepackage{geometry}
\usepackage{natbib}
\usepackage[english]{babel}
\usepackage{graphicx}
\usepackage{epstopdf}
\usepackage{subfigure}
\usepackage{indentfirst}
\usepackage{mathrsfs}
\allowdisplaybreaks[1]

\begin{document}

  \setlength{\parindent}{2em}
  \title{Optical properties of a Brane-World black hole as photons couple to the Weyl tensor}
  \author{He-Xu Zhang} \author{Cong Li} \author{Peng-Zhang He} \author{Qi-Qi Fan} \author{Jian-Bo Deng} \email[Jian-Bo Deng: ]{dengjb@lzu.edu.cn}
  \affiliation{Institute of Theoretical Physics $\&$ Research Center of Gravitation, Lanzhou University, Lanzhou 730000, China}
  \date{\today}

  \begin{abstract}

  In this article, we have investigated the equations of motion of the photons coupled to Weyl tensor by the geometric optics approximation and the corresponding shadow in a Brane-World black hole spacetime. It is shown that there exists a double shadow for a black hole since the coupling photons with different polarization directions propagate along different paths in the spacetime. Furthermore, we discuss the effects of the metric parameter $\alpha$ related to the cosmological constant, X-cold dark matter parameter $\beta$ and the coupling parameter $\lambda$ on the umbra (the overlap region of the double shadow) and the penumbra. We also obtain the finite-distance corrections to the deflection angle of light in the Brane-World black hole spacetime as the photons coupled to Weyl tensor by using a recent geometric method.
  \end{abstract}

  \pacs{~~04.70.Dy, 95.30.Sf, 97.60.Lf}

  \maketitle

\section{Introduction}

  In the context of the unification of the physical forces and also in cosmology, the extra dimensions models have been increasing interest during recent years. Among them the Brane-World cosmological models~\cite{langlois2002brane,brax2003cosmology,maartens2010brane}, motivated by string theory (M-theory), have been proposed in which the standard fields are confined to a four-dimensional (4D) world viewed as a hypersurface (the brane) embedded in a higher-dimensional space-time (the bulk) through which only gravity can propagate~\cite{rubakov2001large,csaki2004tasi}. With the help of the brane scenario, one could possibly solve some disturbing problem of high-energy physics, such as the hierarchy problems (the problem of the big difference between the electroweak scale $M_{EW} \sim 1$ TeV and the Planck scale $M_{pl} \sim 10^{16}$ TeV) and the cosmological constant problem~\cite{arkani1998hierarchy,antoniadis1998new,randall1999large,randall1999alternative,csaki1999cosmology,cline1999cosmological}. As the most well-known model in the Brane-World theory, Randall and Sundrum (RS) models~\cite{randall1999large,randall1999alternative} achieve the confine of the standard fields through the imposition of  $Z_ 2$ symmetry and use of the Israel junction conditions which relates the extrinsic curvature of the brane to the energy-momentum tensor of the matter. However, this method works only when the theories have one extra dimension. In Ref.~\cite{heydari2007brane}, Heydari-Fard and Razmi studied a Brane-World embedded in a m-dimensional bulk by means of the confining potential. The field equations obtained on the brane contained an extra term which was identified with the X-cold dark matter. The same method was used in~\cite{heydari2007anisotropic,heydari2007generalized} to find the spherically symmetric vacuum solutions of the field equations on the brane. These solutions could account for the accelerated expansion of the universe and offered an explanation for the galaxy rotation curves without assuming the existence of dark matter and without working with modified Newtonian dynamics.
\par
  The interactions between electromagnetic field and curvature tensor are not included in the standard Einstein-Maxwell theory. Drummond et al.~\cite{drummond1980qed} first found that such kind of the couplings could be appeared naturally in quantum electrodynamics with the photon effective action originating from one-loop vacuum polarization on a curved background spacetime. The coupling between electromagnetic field and curvature tensor  change the propagation path and dispersion relation of the coupled photons in spacetime£¬which lead to the birefringence of light in  the gravitational field and may result in the superluminal phenomenon in some cases~\cite{daniels1994faster,daniels1996faster}. It is also found that the quantum fluctuation of electromagnetic field caused by the coupling of electromagnetic field and spacetime curvature can cause inflation in the early evolution of the universe~\cite{turner1988inflation,mazzitelli1995scalar,lambiase2004gauge,raya2006gravitoelectromagnetic,campanelli2008inflation}; the fluctuation of the electromagnetic field coupled with the gravitational field can also provide a possible physical mechanism for generating large-scale magnetic fields in the center of the Milky Way~\cite{bamba2008inflation,kim1990halo,kim1991detection,clarke2001new}.
 \par
  Since Weyl tensor is actually related to the curvature tensors $R_ {\mu\nu\rho\sigma}$, the Ricci tensor $R_ {\mu\nu}$ and the Ricci scalar $R$, the theory of electromagnetic field with Weyl corrections can be treated as a special kind of generalized Einstein-Maxwell theory with the coupling between the gravitational and electromagnetic fields. It is shown that the couplings with Weyl tensor change the universal relation with the $U(1)$ central charge in the holographic conductivity in the background of anti-de Sitter spacetime ~\cite{ritz2009weyl} and modify the properties of the holographic superconductor including the critical temperature and the order of the phase transition~\cite{wu20113+,ma2011stuckelberg,momeni2012p,roychowdhury2012effect,momeni2012condensation,momeni2011note,zhao2013holographic}. Moreover, Songbai Chen et al. find that with these couplings the dynamical evolution and Hawking radiation of electromagnetic field in the black hole spacetime depend on the coupling parameter and the parity of the field~\cite{chen2013dynamical,chen2014dynamical,liao2014absorption}.
\par
  Due to the strong gravity and high mass density, the coupling between electromagnetic field and curvature tensor should be emerged reasonably in the region near the classical supermassive compact objects at the center of galaxies. Ni's model has been investigated widely in astrophysics~\cite{solanki2004solar,preuss2004astronomical,itin2003maxwell,dereli2011non} and black hole physics~\cite{balakin2005non,balakin2008nonminimal,hehl2001does}. These investigations show that the coupling term modifies the equations of motion both for electromagnetic and for gravitational fields, which could yield time delays in the arrival of gravitational and electromagnetic waves.
\par
  Recently,  Songbai Chen studied the strong gravitational lensing for the photons coupled to Weyl tensor in Schwarzschild and Keer black hole spacetimes~\cite{chen2015strong,chen2017strong}. And Yang Huang disscussed the double shadow of a regular phantom black hole as photons couple to the Weyl tensor~\cite{huang2016double}. Abbas studied the strong gravitational lensing for photons coupled to Weyl tensor in Kiselev black hole~\cite{abbas2019strong}. In this work, we consider a Brane-World black hole spacetime described by Heydari-Fard and Razmi for which the shadow has not yet been calculated. We obtain the effective metric of the coupled photons by the geometric optics approximation, and then disscuss the black hole shadow. We also calculate the deflection angle of light with finite-distance corrections in the Brane-World black hole spacetime as the photons coupled to Weyl tensor.
\par
  The plan of the paper is as follows. In the next section is the derivation of the equations of motion for the photons coupled to the Weyl tensor in the Brane-World black hole. In Sec.~\uppercase\expandafter{\romannumeral3}, we discuss the Weyl corrections to photon sphere radius and angular radius of the shadow. Sec.~\uppercase\expandafter{\romannumeral4} is the deflection angle of light with finite-distance corrections in the Brane-World black hole. Conclusions and discussions are presented in Sec.~\uppercase\expandafter{\romannumeral5}.

\section{Equations of motion for the photons coupled to the Weyl tensor in the Brane-World black hole}
  In this paper, we begin with the action of the electromagnetic field coupled to Weyl tensor in the curved spacetime, which can be expressed as~\cite{ritz2009weyl}
  \begin{equation}
   \label{eq:action}
   S=\int d^4x\sqrt{-g}\left[\frac{R}{16\pi G}-\frac{1}{4}\left(F_{\mu\nu}F^{\mu\nu}-4\lambda C^{\mu\nu\rho\sigma}F_{\mu\nu}F_{\rho\sigma}\right)\right],
  \end{equation}
  where the electromagnetic tensor $F_{\mu\nu}$ is equal to $F_{\mu\nu}=A_{\mu;\nu}-A_{\nu;\mu}$ and $\lambda$ is the coupling constant with dimension of length-squared.
  Note that $C_{\mu\nu\rho\sigma}$ is the Weyl tensor, which is defined as
  \begin{equation}
   \label{eq:Weyl}
   C_{\mu\nu\rho\sigma}=R_{\mu\nu\rho\sigma}-\left(g_{\mu[\rho}R_{\sigma]\nu}-g_{\nu[\rho}R_{\sigma]\mu}\right)+\frac{1}{3}Rg_{\mu[\rho}g_{\sigma]\nu},
  \end{equation}
  here the brackets around indices refer to the antisymmetric part.
  Varying the action Eq.~(\ref{eq:action}) with respect to $A_{\mu}$, one can obtain the following Maxwell equations with Weyl correction
  \begin{equation}
   \label{eq:correction equation}
   \nabla_{\mu}\left(F_{\mu\nu}-4\lambda C_{\mu\nu\rho\sigma}F^{\rho\sigma}\right)=0.
  \end{equation}
\par
  In order to derive the equations of motion of the photons from the above corrected Maxwell equations, one can adopt to the geometric optics approximation. Under this approximation, the wavelength of photon $\lambda$ is much smaller than a typical curvature scale $L$, but is larger than the electron Compton wavelength $\lambda_{e}$, which ensures that the change of the background gravitational and electromagnetic fields with the typical curvature scale can be neglected for the photon propagation~\cite{drummond1980qed,daniels1994faster,daniels1996faster,shore2002faster,cai1998propagation,cho1997faster,lorenci2000light,dalvit2001one,ahmadi2008quantum}.
  In the geometric optics approximation, we write the electromagnetic field strength as the product of a slowly varying amplitude and a rapidly varying phase, i.e.
  \begin{equation}
   \label{eq:F}
   F_{\mu\nu}=f_{\mu\nu}e^{i\theta},
  \end{equation}
  where the wave vector is $k_{\mu}=\nabla_{\mu}\theta$. In the quantum particle interpretation, we identify it as the photon momentum. The amplitude is constrained by the Bianchi identity to be of the form
  \begin{equation}
   \label{eq:f}
   f_{\mu\nu}=k_{\mu}a_{\nu}-k_{\nu}a_{\mu},
  \end{equation}
  where $a_{\mu}$ is the polarization vector satisfying the condition that $k_{\mu}a^{\mu}=0$. Light rays (photon trajectories) are defined as the integral curves of the wave vector (photon momentum).
  Substituting Eqs.~(\ref{eq:F}) and (\ref{eq:f}) into Eq.~(\ref{eq:correction equation}) and using the relationship above, we can get the equation of motion of photon coupling to the Weyl tensor
  \begin{equation}
   \label{eq:photon motion0}
   k_{\mu}k^{\mu}a^{\nu}+8\lambda C^{\mu\nu\rho\sigma}k_{\sigma}k_{\mu}a_{\rho}=0.
  \end{equation}
  Obviously, the coupling term with Weyl tensor changes the propagation of the coupled photon in the background spacetime.
  It is convenient to introduce the orthonormal frame by using the vierbeins defined as $g_{\mu\nu}=\eta_{ab}e_{\mu}^{a}e_{\nu}^{b}$, where $\eta_{ab}$ is the Minkowski metric. In the orthonormal frames, Eq.~(\ref{eq:photon motion0}) can be rewritten as
  \begin{equation}
   \label{eq:photon motion}
   k^{2}a^{b}+8\lambda C^{abcd}k_{d}k_{a}a_{c}=0.
  \end{equation}
\par
  As mentioned above, Heydari-Fard and Razmi have gotten a black hole solution with X-cold dark matter in Brane-World theory described by the spacetime metric ~\cite{heydari2007brane}
  \begin{equation}
   \label{eq:metric}
   ds^2=-f\left(r\right)dt^2+\frac{dr^2}{f\left(r\right)}+r^2\left(d\theta^2+sin^2\theta d\phi^2\right),
  \end{equation}
  with
  \begin{equation}
   \label{eq:g00}
   f\left(r\right)=1-\frac{2M}{r}-\alpha^2 r^2-2\alpha\beta r-\beta^2.
  \end{equation}
  Here, $\alpha$ is a metric parameter which is related to the cosmological constant and $\beta$ is a parameter of X-cold dark matter. It is the Schwarzchild-de Sitter-like solution as $\beta =0$ and the Schwarzschild-XCDM solution as $\alpha =0$.
\par
  We now introduce a local orthonormal frame. The appropriate basis l-forms are $e^a\left(a=0,1,2,3\right)$ with
  \begin{equation}
   \label{eq:frame}
   e^0=\sqrt{f}dt,\qquad e^1=\frac{1}{\sqrt{f}}dr,\qquad e^2=rd\theta,\qquad e^3=rsin\theta d\phi.
  \end{equation}
  By a straightforward calculation, in the orthonormal frame the independent nonvanishing components of the Riemann curvature tensor are
  \begin{equation}
   \begin{aligned}
   \label{eq:R}
   &R_{0101}=-\frac{2M}{r^3}-\alpha^2, & R_{0202}=R_{0303}=\frac{M}{r^3}-\frac{\alpha\beta}{r}-\alpha^2,\\
   &R_{1212}=R_{1313}=-\frac{M}{r^3}+\frac{\alpha\beta}{r}+\alpha^2, & R_{2323}=\frac{2M}{r^3}+\frac{\beta^2}{r^2}+\frac{2\alpha\beta}{r}+\alpha^2.
   \end{aligned}
  \end{equation}
  Introducing the notation $U_{ab}^{01}\equiv \delta_{a}^{0}\delta_{b}^{1}-\delta_{a}^{1}\delta_{b}^{0}$, etc., one can rewrite the complete Weyl tensor compactly in the following form
  \begin{equation}
   \label{eq:Weyl 2}
   C_{abcd}=\mathcal{A}\left(2U_{ab}^{01}U_{cd}^{01}-U_{ab}^{02}U_{cd}^{02}-U_{ab}^{03}U_{cd}^{03}+U_{ab}^{12}U_{cd}^{12}+U_{ab}^{13}U_{cd}^{13}-2U_{ab}^{23}U_{cd}^{23}\right),
  \end{equation}
  with
  \begin{equation}
   \label{eq:A}
   \mathcal{A}=-\left(\frac{M}{r^3}+\frac{\beta^2}{6r^2}\right).
  \end{equation}
  In order to solve the equation of motion for the coupled photon propagation, one can introduce three linear combinations of momentum components~\cite{daniels1996faster,cai1998propagation}
  \begin{equation}
   \label{eq:l}
   l_{b}=k^{a}U_{ab}^{01},\qquad m_{b}=k^{a}U_{ab}^{02},\qquad n_{b}=k^{a}U_{ab}^{03},
  \end{equation}
  and some dependent combinations
  \begin{equation}
   \label{eq:p}
   p_{b}=k^{a}U_{ab}^{12},\qquad q_{b}=k^{a}U_{ab}^{13},\qquad r_{b}=k^{a}U_{ab}^{23}.
  \end{equation}
  Using $l_{b}$, $m_{b}$, and $n_{b}$ to contract Eq.~(\ref{eq:photon motion}), with the help of Eq.~(\ref{eq:Weyl 2}), the photon equations of motion coupling with Weyl tensor can easily be simplified as a set of equations for three independent polarisation components $a\cdot l$, $a\cdot m$ and $a\cdot r$
  \begin{equation}
   \label{eq:matrix}
   \begin{pmatrix}K_{11}&0&0\\K_{21}&K_{22}&K_{23}\\0&0&K_{33}\end{pmatrix}\begin{pmatrix}a\cdot l\\a\cdot m\\a\cdot r\end{pmatrix}=0,
  \end{equation}
  with
  \begin{equation}
   \begin{split}
   \label{eq:K}
   &K_{11}=\left(1-16\lambda\mathcal{A}\right)\left(-k^{0}k^{0}+k^{1}k^{1}\right)+\left(1+8\lambda\mathcal{A}\right)\left(k^{2}k^{2}+k^{3}k^{3}\right),\\
   &K_{21}=-24\lambda\mathcal{A}k^{1}k^{2},\\
   &K_{22}=\left(1+8\lambda\mathcal{A}\right)\left(-k^{0}k^{0}+k^{1}k^{1}+k^{2}k^{2}+k^{3}k^{3}\right),\\
   &K_{23}=24\lambda\mathcal{A}k^{0}k^{3},\\
   &K_{33}=\left(1+8\lambda\mathcal{A}\right)\left(-k^{0}k^{0}+k^{1}k^{1}\right)+\left(1-16\lambda\mathcal{A}\right)\left(k^{2}k^{2}+k^{3}k^{3}\right).
   \end{split}
  \end{equation}
  The condition of Eq.~(\ref{eq:matrix}) with non-zero solution is $K_{11}K_{22}K_{33}=0$.
  The first root $K_{11}=0$ leads to the modified light cone
  \begin{equation}
   \label{eq:K11}
   \left(1-16\lambda\mathcal{A}\right)\left(-k^{0}k^{0}+k^{1}k^{1}\right)+\left(1+8\lambda\mathcal{A}\right)\left(k^{2}k^{2}+k^{3}k^{3}\right)=0,
  \end{equation}
  which corresponds to the case the polarisation vector $a_{\mu}$ is proportional to $l_{\nu}$ and the strength $f_{\mu\nu}\varpropto\left(k_{\mu}l_{\nu}-k_{\nu}l_{\mu}\right)$.
  The second root $K_{22}=0$ means that $a\cdot l=0$ and $a\cdot r=0$ in Eq.~(\ref{eq:matrix}), which implies $a_{\mu}=\lambda k_{\mu}$ and then $f_{\mu\nu}$ vanishes~\cite{drummond1980qed}. Thus, this root corresponds to an unphysical polarisation and it should be neglected for general directions of propagation of the coupled photon.
  The third root is $K_{33}=0$, i.e.,
  \begin{equation}
   \label{eq:K33}
   \left(1+8\lambda\mathcal{A}\right)\left(-k^{0}k^{0}+k^{1}k^{1}\right)+\left(1-16\lambda\mathcal{A}\right)\left(k^{2}k^{2}+k^{3}k^{3}\right)=0,
  \end{equation}
  which means that the vector $a_{\mu}=\lambda r_{\mu}$ and the strength $f_{\mu\nu}\varpropto\left(k_{\mu}r_{\nu}-k_{\nu}r_{\mu}\right)$.
\par
  It is easy to find that the light cone conditions Eq.~(\ref{eq:K11}) and Eq.~(\ref{eq:K33}) indicate that the motion of the coupled photons in the equatorial plane is non-geodesic in the Kerr metric. In fact, these
  photons follow null geodesics of the effective metric $\gamma_{\mu\nu}$, i.e., $\gamma^{\mu\nu}k_{\mu}k_{\nu}=0$~\cite{preuss2004astronomical}.
  Moreover, the effects of Weyl tensor on the photon propagation are different for the coupled photons with different polarizations, which yields a phenomenon of birefringence in the spacetime~\cite{daniels1994faster,daniels1996faster,shore2002faster,cai1998propagation,cho1997faster,de2000light,de2000light}.
\par
  The effective metric for the coupled photon in the standard Boyer-Lindquist coordinates can be expressed as
  \begin{equation}
   \label{eq:effective metric}
   ds^2=-f\left(r\right)dt^2+\frac{dr^2}{f\left(r\right)}+W\left(r\right)^{-1} r^2\left(d\theta^2+sin^2\theta d\phi^2\right).
  \end{equation}
  The quantity $W\left( r\right)$ is
  \begin{equation}
   \label{eq:W1}
   W\left(r\right)=\frac{6r^3-8\lambda\left(6M+\beta^{2} r\right)}{6r^3+16\lambda\left(6M+\beta^{2} r\right)},
  \end{equation}
  for photon with the polarization along $l_{\mu}$ (PPL) and
  \begin{equation}
   \label{eq:W2}
   W\left(r\right)=\frac{6r^3+16\lambda\left(6M+\beta^{2} r\right)}{6r^3-8\lambda\left(6M+\beta^{2} r\right)},
  \end{equation}
  for photon with the polarization along $r_{\mu}$ (PPR), respectively.
\section{Weyl corrections to photon sphere radius and angular radius of the shadow}
\subsection{Photon sphere radius}
  In this subsection, we will discuss the Weyl correction to photon sphere radius and angular radius of the shadow. For simplicity, we here just consider that the whole trajectory of the photon is limited on the equatorial plane $\theta =\frac{\pi}{2}$.
  Since the existence of cyclic coordinates $t$ and $\phi$ in spacetime Eq.~(\ref{eq:effective metric}), we can obtain two constants of motion
  \begin{equation}
   \label{eq:Con}
   E=f\left(r\right)\dot{t},\qquad L=r^2W\left(r\right)^{-1}\dot{\phi},
  \end{equation}
  where a dot represents a derivative with respect to affine parameter $\lambda$ along the geodesics. $E$ and $L$ are, respectively, the energy and angular momentum of the photon.
\par
  Using the the null geodesics of the effective metric condition $\gamma^{\mu\nu}k_{\mu}k_{\nu}=0$, one can find that
  \begin{equation}
   \label{eq:D}
   \left(\frac{\mathrm{d}r}{\mathrm{d}\lambda}\right)^2=f\left(r\right)\left[\frac{E^2}{f\left(r\right)}-W\left(r\right)\frac{L^2}{r^2}\right].
  \end{equation}
  Combining Eq.~(\ref{eq:D}) with $\frac{\mathrm{d}r}{\mathrm{d}\lambda}=0$ and $\frac{\mathrm{d^2}r}{\mathrm{d}\lambda^2}=0$, we obtain the photon sphere radius $r_{ph}$ in the equatorial plane satisfying the condition
  \begin{equation}
   \label{eq:circle}
   W\left(r\right)\left[rf'\left(r\right)-2f\left(r\right)\right]+rW'\left(r\right)f\left(r\right)=0.
  \end{equation}
  First, as an example, we show the effect of parameter $\alpha$ related to the cosmological constant on the photon sphere radius $r_{ph}$ and the angle radius $\theta_{sh}$ of the black hole shadow (the approach is given in the next subsection) for PPL in Table~\ref{tab:notation}. According to Table~\ref{tab:notation}, it is easy to find that the difference of $\theta_{sh}$ between $\alpha=10^{-12}$ and $\alpha=10^{-20}$ is of order $10^{-3}\sim10^{-2} ~\mu as$, which far exceeds the angular resolution of the Event Horizon Telescope (EHT) and the space-based very-long baseline interferometry (VLBI) RadioAstron~\cite{doeleman2008event,kardashev2013radioastron}. Therefore we only present the variation of $r_{ph}$ with the coupling factor $\alpha$ and X-cold dark matter parameter $\beta$ for PPL and PPR in Figs.~1 and 2, respectively.
  \begin{table}[h]
   \centering
   \begin{tabular}{|c|c|c|c|c|c|}
		\hline
		$\alpha $ & $\quad 10^{-12}$ & $\quad 10^{-14}$ & $\quad 10^{-16}$ & $\quad 10^{-18}$ & $\quad 10^{-20}$ \\
		\hline
        $r_{ph} $ & 1.639032 & 1.639032 & 1.639032 & 1.639032 & 1.639032 \\
        \hline
		$\theta_{sh}(\mu as) $ & 29.2310 & 29.2931 & 29.2937 & 29.2937 & 29.2937 \\
		\hline
   \end{tabular}
   \caption{~~Dependence of $r_{ph}$ and the angular radius of the shadow $\theta_{sh}$ on the parameter $\alpha$ related to the cosmological constant for PPL. Here we choose Sgr A* as an example and set 2M = 1, $\beta=0.1$, and $\lambda=0.05$.}
   \label{tab:notation}
  \end{table}
  \begin{figure}[htbp]
   \centering
   \includegraphics[width=6.4cm]{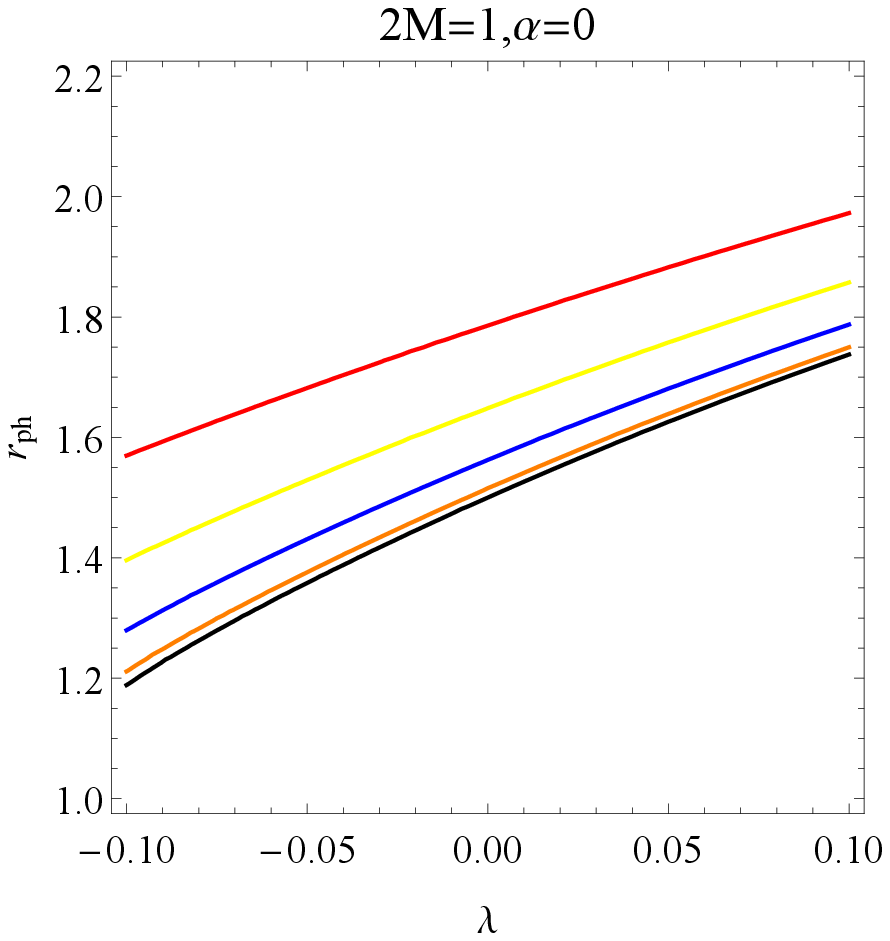}
   \includegraphics[width=2cm]{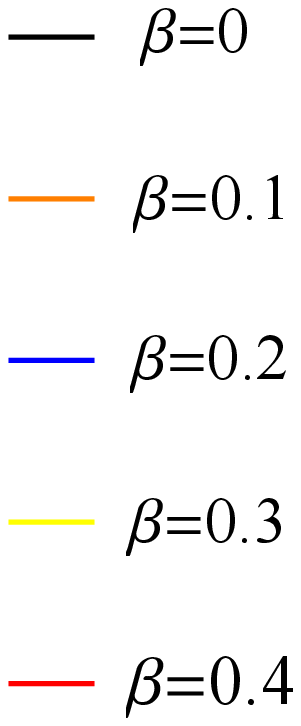}
   \includegraphics[width=6.4cm]{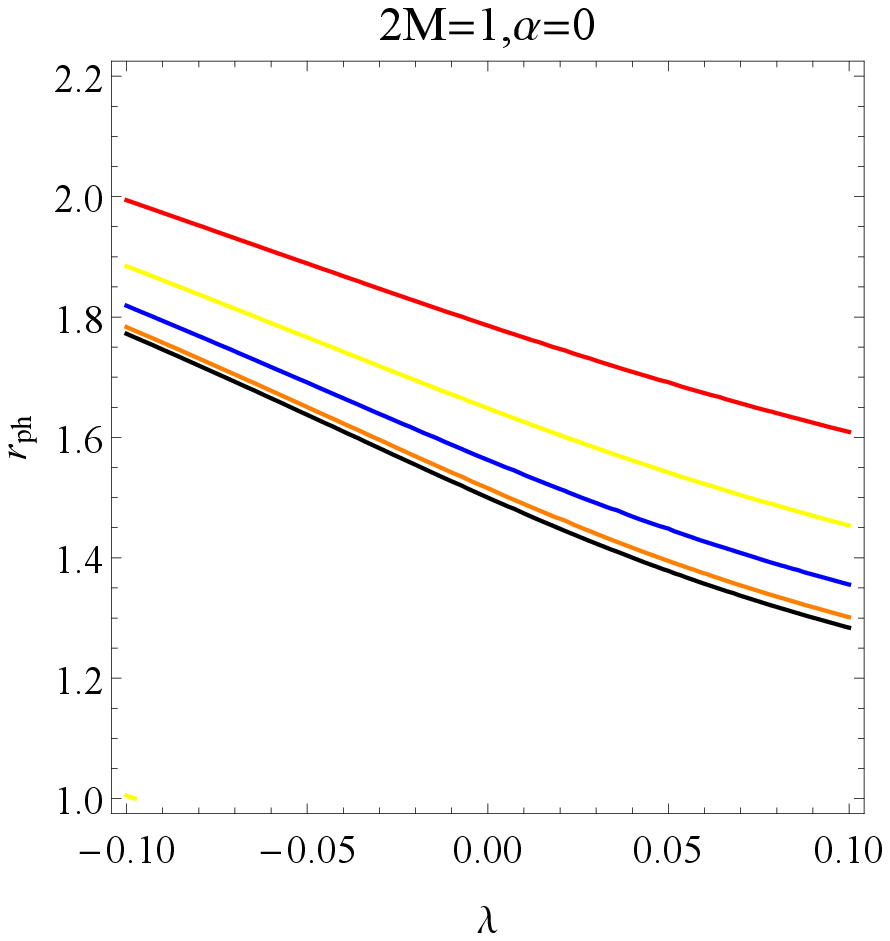}
   \caption{~~Variety of the photon sphere radius $r_{ph}$ with the coupling constant $\lambda$ in a Brane-World black hole spacetime. The left and the right are for PPL and PPR, respectively.}
  \end{figure}
  \begin{figure}[htbp]
   \centering
   \includegraphics[width=6.4cm]{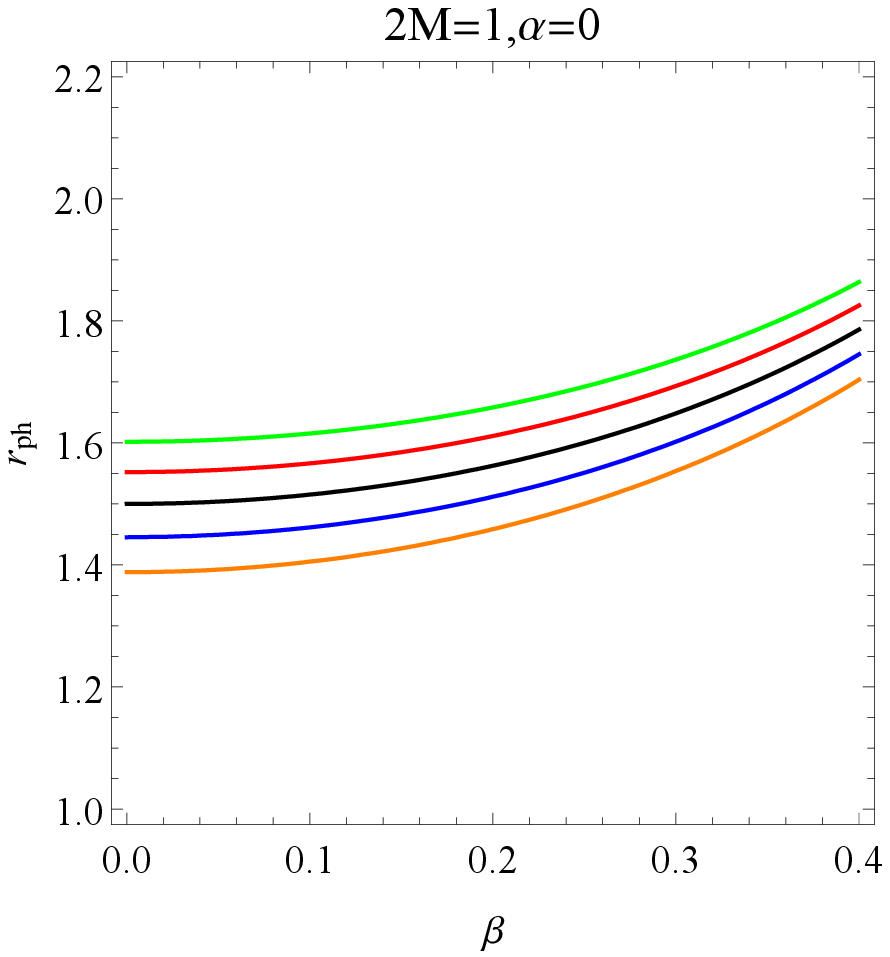}
   \includegraphics[width=2cm]{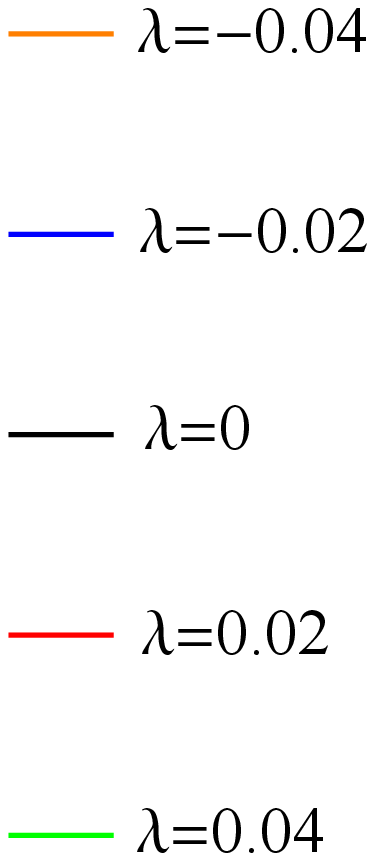}
   \includegraphics[width=6.4cm]{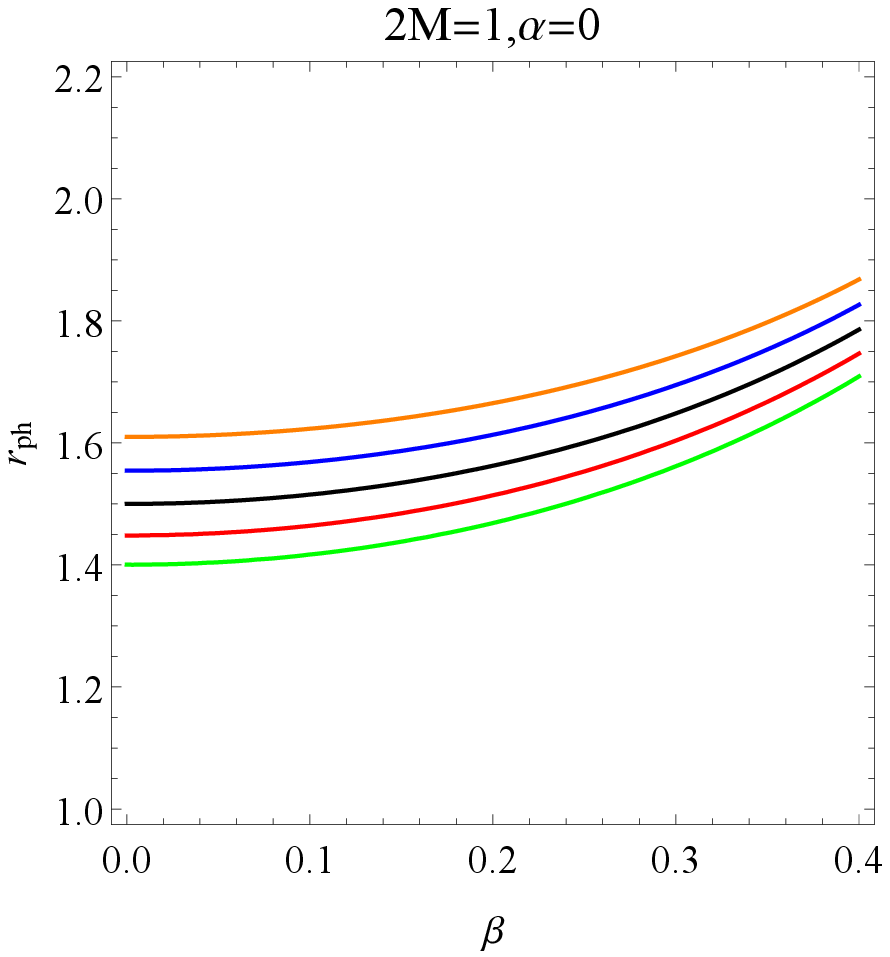}
   \caption{~~Variety of the photon sphere radius $r_{ph}$ with the X-cold dark matter parameter $\beta$ in a Brane-World black hole spacetime. The left and the right are for PPL and PPR, respectively.}
  \end{figure}
\par
  It is shown that, for different values of $\beta$, the photon sphere radius $r_{ph}$  increases with the coupling parameter $\lambda$ for PPL and decreases for PPR. Moreover, with the increase of $\beta$, $r_{ph}$ for a fixed value of the coupling parameter $\lambda$ increases for two different coupled photons. In general, the photon sphere radius $r_{ph}$ depends on the coupling parameter $\lambda$, $\beta$, and the polarization, which is quite different from that in the case without the coupling. In other words, the presence of the coupling leads to the diversity of the photon sphere radius $r_{ph}$.
\subsection{Angular radius of the shadow}
  For the simplest case of a non-rotating black hole, the shadow is a circular disk in the sky. If the black hole is uncharged, it is to be modeled by the Schwarzschild metric. For a static observer in the spacetime of a Schwarzschild black hole, the angular radius of the shadow was calculated in a seminal paper by Synge~\cite{synge1966escape}.
  Making use of this method, one can obtain the corrected angular radius of the shadow $\theta_{sh}$ in the Brane-World black hole spacetime (2)
  \begin{equation}
   \label{eq:angular}
   sin^2 \theta_{sh}=\frac{f\left(r_{O}\right)r_{ph}^2W\left(r_{O}\right)}{f\left(r_{ph}\right)r_{O}^2W\left(r_{ph}\right)}.
  \end{equation}
  Here, we assume that the observer is located at radius coordinate $r_{O}$ with angular coordinate $\theta_{O}=\frac{\pi}{2}$.
\par
  As an example, we consider the supermassive black hole Sgr A* located at the Galactic center. The mass is estimated to be $M$ = 4.4¡Á$\times 10^{6}$ M$_{\odot}$ and its distance from the earth is around 8.5~kpc~\cite{genzel2010galactic}. Substituting these data into the Eq.~(\ref{eq:angular}), the angular radius of the shadow yielded by the photons coupling with the Weyl tensor in the Brane-World spacetime are obtained and shown in Figs.~3 and 4.
  \begin{figure}[htbp]
   \centering
   \includegraphics[width=6.4cm]{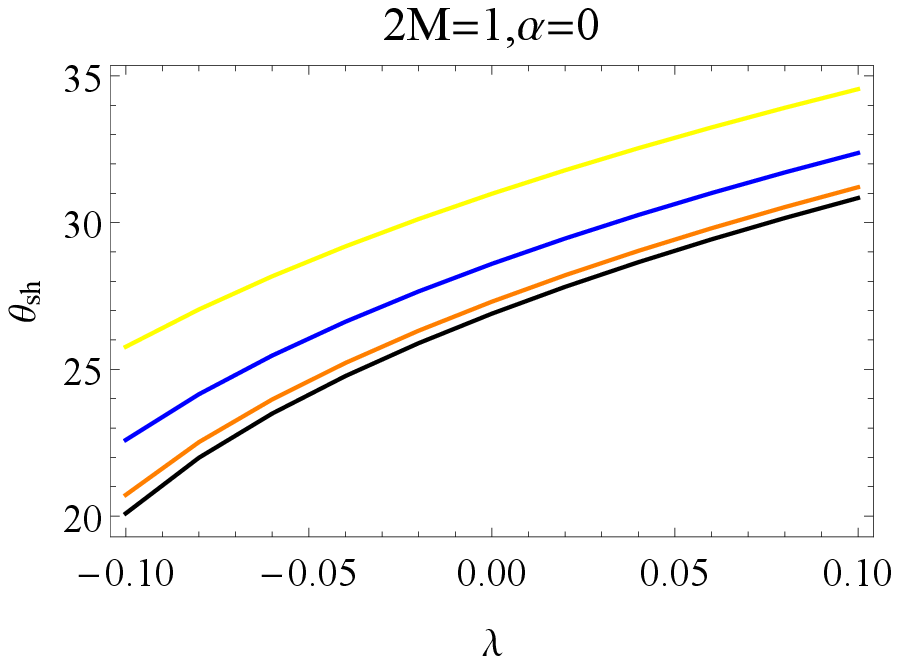}
   \includegraphics[width=2cm]{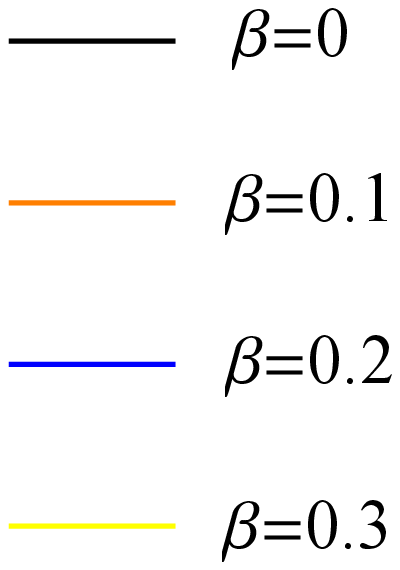}
   \includegraphics[width=6.4cm]{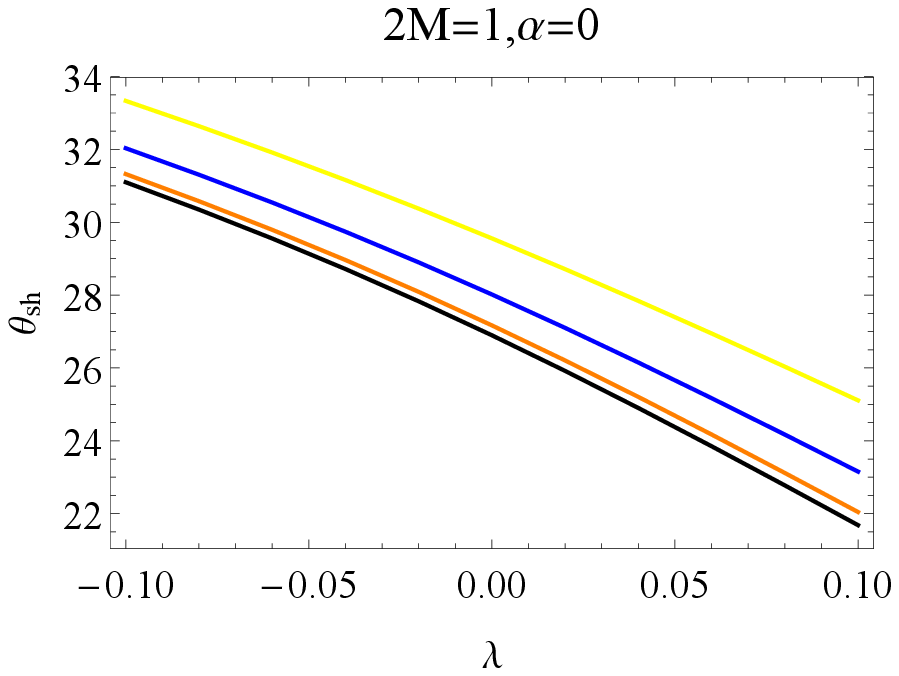}
   \caption{~~Variety of the angular radius of the shadow with the coupling constant $\lambda$ in a Brane-World black hole spacetime. The left and the right are for PPL and PPR, respectively.}
  \end{figure}
  \begin{figure}[htbp]
   \centering
   \includegraphics[width=6.4cm]{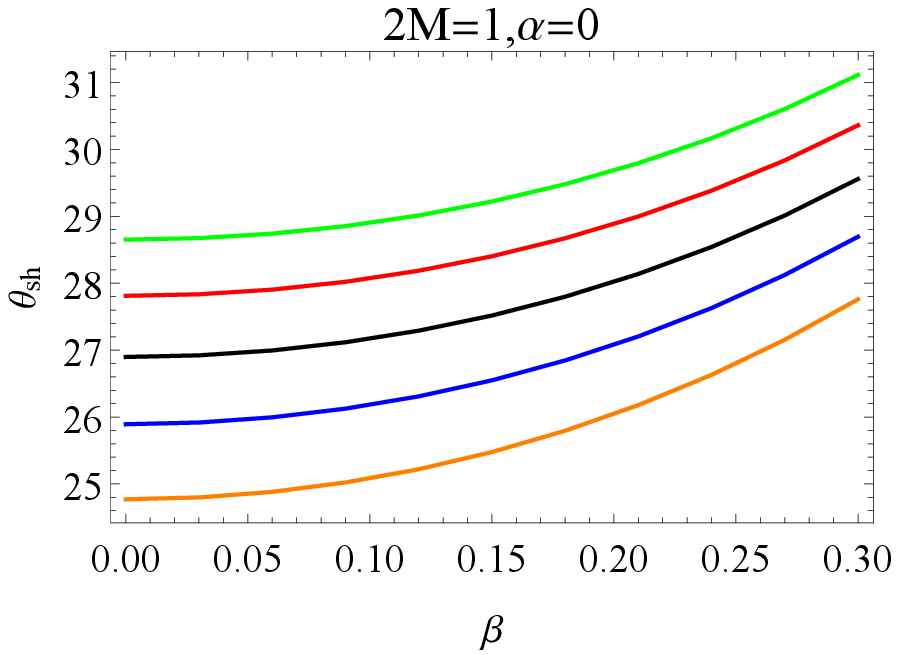}
   \includegraphics[width=2cm]{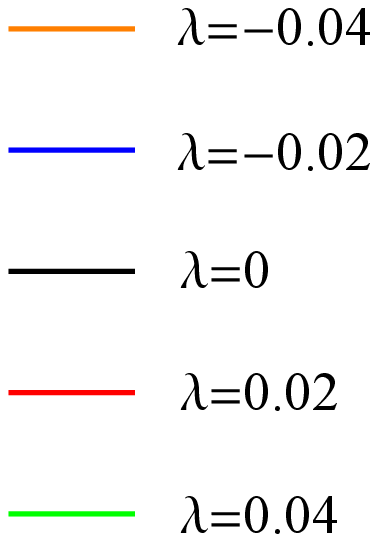}
   \includegraphics[width=6.4cm]{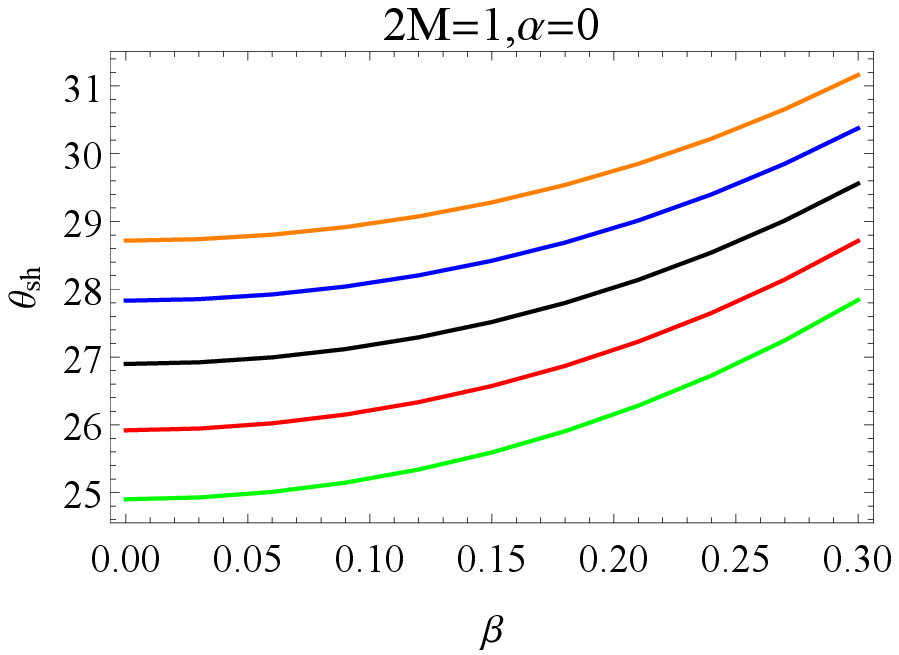}
   \caption{~~Variety of the angular radius of the shadow with the X-cold dark matter parameter $\beta$ in a Brane-World black hole spacetime. The left and the right are for PPL and PPR, respectively.}
  \end{figure}
\par
  From Fig.~3, we find that for different $\beta$, the angular radius of the shadow $\theta_{sh}$ monotonically increases with the coupling parameter $\lambda$ for PPL and decreases for PPR. In Fig.~4, with the increase of $\beta$, the angular radius of the shadow $\alpha_{sh}$  increases for two different coupled photons.
  Considering that due to the coupling photons with different polarization directions propagate along different paths in the spacetime, it is naturally expected that there exists a double shadow for a black hole as photons couple to the Weyl tensor. The overlap region of the double shadow is called an umbra. In Fig.~3, we see that, for different parameter $\beta$, the size of umbra is always determined by PPL when $\lambda$ is negative and by PPR when $\lambda$ is positive. Moreover, one can find that the umbra of the black hole increases with the X-cold dark matter parameter $\beta$ and decreases with the coupling strength. And they coincide with the conclusions in Ref.~\cite{huang2016double}.
  In order to see more clearly the effect of $\beta$ and the coupling parameter $\lambda$ on the angular radius of penumbra of the black hole $\Delta\theta_{sh}$, we have made Fig.~5. It is shown that the size of penumbra increases with the coupling parameter $\lambda$ and decreases with the X-cold dark matter parameter $\beta$. We can find that $\Delta\theta_{sh}=0$ without the coupling, which means the double shadow of the black hole is reduced to a single shadow.
  \begin{figure}[htbp]
   \centering
   \includegraphics[width=6.4cm]{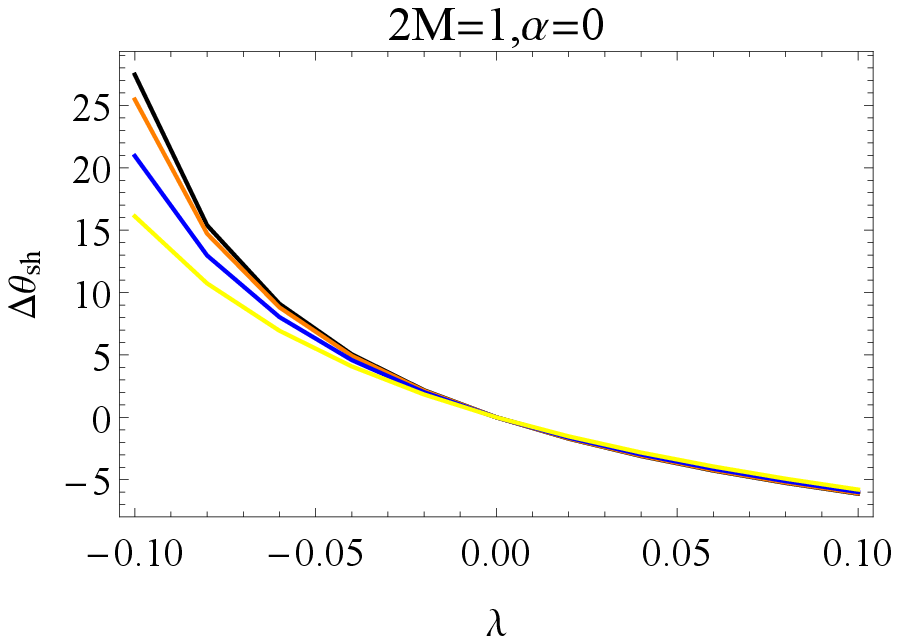}
   \includegraphics[width=2cm]{c.eps}
   \includegraphics[width=6.4cm]{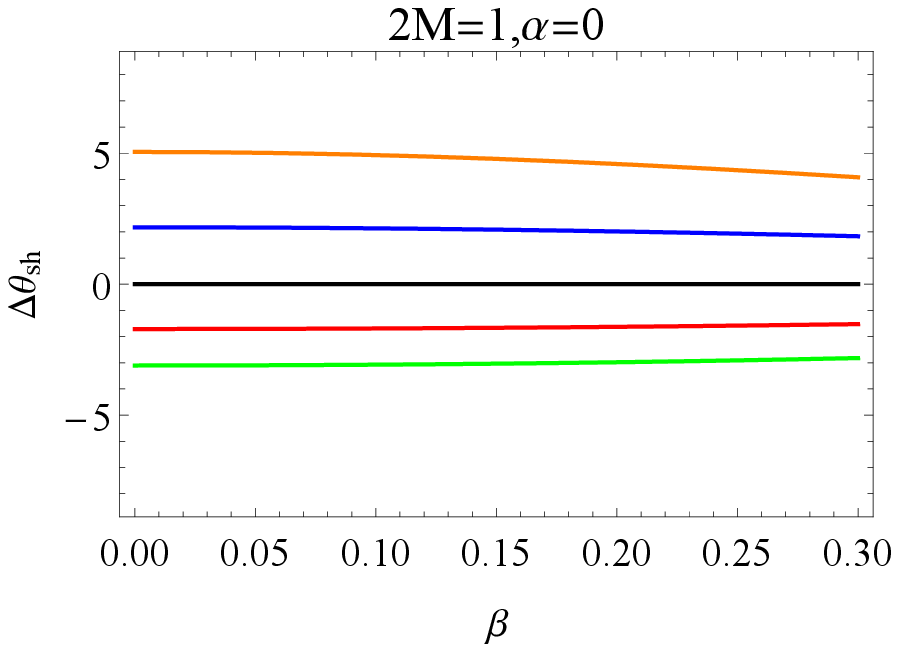}
   \caption{~~Evolution of the penumbra of the black hole with the coupling constant $\lambda$ and the X-cold dark matter parameter $\beta$ in the Brane-World black hole spacetime.}
  \end{figure}
\section{Deflection angle of light with finite-distance corrections in the Brane-World black hole}
  In this section, we proceed to study the deflection angle of light using the  methodology given by Ishihara et al.~\cite{ishihara2016gravitational,ono2017gravitomagnetic} under the assumption that the distance from the source (S) to the receiver (R) is finite.
\par
  Light rays satisfy the null condition as $ds^2=0$, which is rearranged as, via Eq.~(\ref{eq:effective metric})
  \begin{equation}
   \label{eq:optical metric}
   dt^2=\widetilde{\gamma}_{ij}dx^{i}dx^{j}=\frac{dr^2}{f^2\left(r\right)}+\frac{r^2}{W\left(r\right)f\left(r\right)},
  \end{equation}
  where $i$ and $j$ denote 1 and 2, $\widetilde{\gamma}_{ij}$ is called the optical metric. The optical metric defines a 3-dimensional Riemannian space (denoted as $M^{opt}$), in which the light ray is expressed as a spatial curve.
  As usual, we define the impact parameter of the light ray as
  \begin{equation}
   \label{eq:b}
   b\equiv\frac{L}{E}=\frac{r^2}{W\left(r\right)f\left(r\right)}\frac{d\phi}{dt}.
  \end{equation}
  By using  Eqs.~(\ref{eq:optical metric}) and (\ref{eq:b}) by introducing a new variable as $u=r^{-1}$, one can obtain the orbit equation as
  \begin{equation}
   \label{eq:u}
   \left(\frac{du}{d\phi}\right)^2=\frac{1}{b^2W^2}-\frac{fu^2}{W}\equiv F\left(u\right).
  \end{equation}
  Then, $\phi_{RS}$ is obtained as
  \begin{equation}
   \begin{split}
   \label{eq:jifenjiaoPPL}
   \phi_{RS}=&\int_{S}^{R} d\phi=\left[\pi-\arcsin(bu_{R})-\arcsin(bu_{S})\right]+\frac{M}{b}\left(\frac{2-b^2u_{R}^2}{\sqrt{1-b^2u_{R}^2}}+\frac{2-b^2u_{S}^2}{\sqrt{1-b^2u_{S}^2}}\right)\\
   &+\frac{8M\lambda}{b^3}\left(\frac{2-b^2u_{R}^2}{\sqrt{1-b^2u_{R}^2}}+\frac{2-b^2u_{S}^2}{\sqrt{1-b^2u_{S}^2}}\right)+\phi_{1}^{\ast}\left(\alpha,\beta,\lambda\right)\\
   &+\mathcal{O}\left(bM^2u_{R}^3,bM^2u_{S}^3,\frac{M^2\lambda u_{R}^3}{b},\frac{M^2\lambda u_{S}^3}{b},\frac{M^2\lambda^2 u_{R}^3}{b^3},\frac{M^2\lambda^2 u_{S}^3}{b^3},b^3\alpha^2 u_{R},b^3{\alpha^2 u_{S}}\right),
   \end{split}
  \end{equation}
  for PPL and
  \begin{equation}
   \begin{split}
   \label{eq:jifenjiaoPPM}
   \phi_{RS}=&\int_{S}^{R} d\phi=\left[\pi-\arcsin(bu_{R})-\arcsin(bu_{S})\right]+\frac{M}{b}\left(\frac{2-b^2u_{R}^2}{\sqrt{1-b^2u_{R}^2}}+\frac{2-b^2u_{S}^2}{\sqrt{1-b^2u_{S}^2}}\right)\\
   &-\frac{8M\lambda}{b^3}\left(\frac{2-b^2u_{R}^2}{\sqrt{1-b^2u_{R}^2}}+\frac{2-b^2u_{S}^2}{\sqrt{1-b^2u_{S}^2}}\right)+\phi_{2}^{\ast}\left(\alpha,\beta,\lambda\right)\\
   &+\mathcal{O}\left(bM^2u_{R}^3,bM^2u_{S}^3,\frac{M^2\lambda u_{R}^3}{b},\frac{M^2\lambda u_{S}^3}{b},\frac{M^2\lambda^2 u_{R}^3}{b^3},\frac{M^2\lambda^2 u_{S}^3}{b^3},b^3\alpha^2 u_{R},b^3{\alpha^2 u_{S}}\right),
   \end{split}
  \end{equation}
  for PPR, respectively. Here, the detailed forms of $\phi_{1}^{\ast}$ and $\phi_{2}^{\ast}$ are shown in the appendix.
\par
  Let $\Psi$ denote the angle of the light ray measured from the radial direction. Accordingly, the $\Psi_{R}$ and $\Psi_{S}$ denote the angles that are measured at the receiver position ($R$) and the source position ($S$), respectively. Please see Fig.~6.\\
  \begin{figure}[htbp]
   \centering
   \includegraphics[width=10cm]{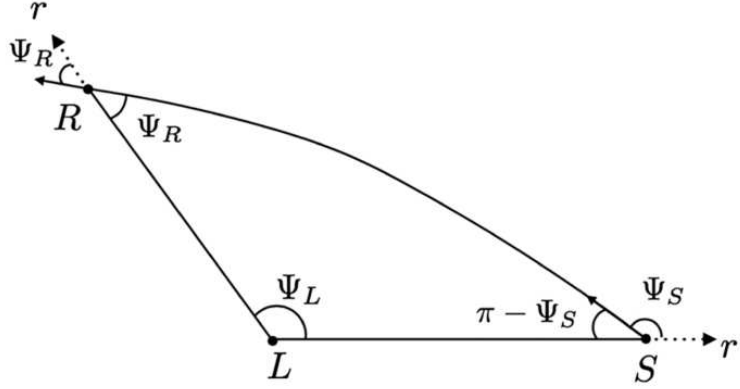}
   \caption{~~ Triangle embedded in the curved space $M^{opt}$. Here, the point $L$ denotes the lens center.}
  \end{figure}\\
  Using the unit tangent vector $e_{i}$ along the light ray orbit in the space $M^{opt}$ which satisfies the relation $\widetilde{\gamma}_{ij}e^{i}e^{j}=1$, we can obtain $\Psi$ in terms of the following relation~\cite{ishihara2016gravitational}
  \begin{equation}
   \label{eq:psi}
   \sin\Psi=\frac{b\sqrt{Wf}}{r}.
  \end{equation}
  By Eq.~(\ref{eq:psi}), we find that $\Psi_{R}-\Psi_{S}$ for both PPL and PPR take this form under the low-order approximations:
  \begin{equation}
   \begin{split}
   \label{eq:deflect angular PPL}
   \Psi_{R}-\Psi_{S}=&\Phi_{R}^{Sch}-\Phi_{S}^{Sch}-\frac{b}{\sqrt{1-b^{2}u_{R}^{2}}}\left[\alpha\beta+\frac{\alpha^{2}}{2u_{R}}+\frac{\alpha^2M}{2\left(1-b^{2}u_{R}^{2}\right)}\right]\\
   &-\frac{b}{\sqrt{1-b^{2}u_{S}^{2}}}\left[\alpha\beta+\frac{\alpha^{2}}{2u_{S}}+\frac{\alpha^2M}{2\left(1-b^{2}u_{S}^{2}\right)}\right]\\
   &+\mathcal {O}\left(bM^2 u_{R}^3,bM^2 u_{S}^3,bM^2\alpha^2 u_{R},bM^2\alpha^2 u_{S},bM\alpha\beta u_{R},bM\alpha\beta u_{S},b\beta^2 u_{R},b\beta^2 u_{S}\right),
   \end{split}
  \end{equation}
  where
  \begin{equation}
   \label{eq:Sch}
   \Phi_{R}^{Sch}-\Phi_{S}^{Sch}\equiv \left[\arcsin\left(bu_{R}\right)+\arcsin\left(bu_{S}\right)-\pi\right]-bM\left(\frac{u_{R}^{2}}{\sqrt{1-b^{2}u_{R}^{2}}}+\frac{u_{S}^{2}}{\sqrt{1-b^{2}u_{S}^{2}}}\right).
  \end{equation}
  Let $\phi_{RS}\equiv\phi_{R}-\phi_{S}$ denote the coordinate separation angle between the receiver and source.
  Then the deflection angle $\hat{\alpha}$ is expressed as~\cite{ishihara2016gravitational}
  \begin{equation}
   \label{eq:alpha}
   \hat{\alpha}\equiv\Psi_{R}-\Phi_{S}+\phi_{RS},
  \end{equation}
  where the closest approach $r_{0}=u_{0}^{-1}$.
  That is to say, basically, one can find the deflection angle $\hat{\alpha}$ by just computing $\Psi_{R}$, $\Psi_{S}$, and $\phi_{RS}$ and applying Eq.~(\ref{eq:alpha}).
\par
  Inserting Eqs.~(\ref{eq:jifenjiaoPPL}), (\ref{eq:jifenjiaoPPM}) and (\ref{eq:deflect angular PPL}) into Eq.~(\ref{eq:alpha}), the deflection angle can be obtained as
  \begin{equation}
   \begin{split}
   \label{eq:alpha hat PPL}
   \hat{\alpha}=&\frac{M}{b}\left(\frac{2-b^2u_{R}^2}{\sqrt{1-b^2u_{R}^2}}+\frac{2-b^2u_{S}^2}{\sqrt{1-b^2u_{S}^2}}\right)+\frac{8M\lambda}{b^3}\left(\frac{2-b^2u_{R}^2}{\sqrt{1-b^2u_{R}^2}}+\frac{2-b^2u_{S}^2}{\sqrt{1-b^2u_{S}^2}}\right)\\
   &-M\left(\frac{bu_{R}^2}{\sqrt{1-b^2u_{R}^2}}+\frac{bu_{S}^2}{\sqrt{1-b^2u_{S}^2}}\right)-\frac{b\alpha^2}{2}\left(\frac{1}{u_{R}\sqrt{1-b^2u_{R}^2}}+\frac{1}{u_{S}\sqrt{1-b^2u_{S}^2}}\right)+\hat{\alpha}_{1}^{\ast}\\
   &+\mathcal {O}\left(bM^2u_{R}^3,bM^2u_{S}^3,\frac{M^2\lambda u_{R}^3}{b},\frac{M^2\lambda u_{S}^3}{b},\cdots,b\beta^2 u_{R},b\beta^2 u_{S}\right),
   \end{split}
  \end{equation}
  for PPL.
  Similarly, for PPR, we have
  \begin{equation}
   \begin{split}
   \label{eq:alpha hat PPM}
   \hat{\alpha}=&\frac{M}{b}\left(\frac{2-b^2u_{R}^2}{\sqrt{1-b^2u_{R}^2}}+\frac{2-b^2u_{S}^2}{\sqrt{1-b^2u_{S}^2}}\right)-\frac{8M\lambda}{b^3}\left(\frac{2-b^2u_{R}^2}{\sqrt{1-b^2u_{R}^2}}+\frac{2-b^2u_{S}^2}{\sqrt{1-b^2u_{S}^2}}\right)\\
   &-M\left(\frac{bu_{R}^2}{\sqrt{1-b^2u_{R}^2}}+\frac{bu_{S}^2}{\sqrt{1-b^2u_{S}^2}}\right)-\frac{b\alpha^2}{2}\left(\frac{1}{u_{R}\sqrt{1-b^2u_{R}^2}}+\frac{1}{u_{S}\sqrt{1-b^2u_{S}^2}}\right)+\hat{\alpha}_{2}^{\ast}\\
   &+\mathcal {O}\left(bM^2 u_{R}^3,bM^2 u_{S}^3,bM^2\alpha^2 u_{R},bM^2\alpha^2 u_{S},bM\alpha\beta u_{R},bM\alpha\beta u_{S},b\beta^2 u_{R},b\beta^2 u_{S}\right).
   \end{split}
  \end{equation}
  Here, the detailed forms of $\hat{\alpha}_{1}^{\ast}$ and $\hat{\alpha}_{2}^{\ast}$ are shown in the appendix.
\par
  Note that some terms in these two expressions may apparently diverge as $bu_{R}\rightarrow 0$ and $bu_{S}\rightarrow 0$. This is because the spacetime is not asymptotically flat and hence it does not allow the limit of $bu_{R}\rightarrow 0$ and $bu_{S}\rightarrow 0$. However, from a physical point of view, we know that an observed star or galaxy is located at a finite distance from us~\cite{ishihara2016gravitational,haroon2019shadow}. In other words, we can consider only finitedistance corrections in this case, one can include only a certain finite distance which leads to the further simplified relations
  \begin{equation}
   \begin{split}
   \label{eq:M1}
   \hat{\alpha}\sim&\frac{4M}{b}-\frac{b\alpha^2}{2}\left(\frac{1}{u_{R}}+\frac{1}{u_{S}}\right)+bM\alpha^2+\frac{32M\lambda}{b^3}+\frac{48M^2\alpha\beta}{b}+\frac{8M\beta^2}{b}+\frac{768M^2\alpha\beta\lambda}{b^3}\\
   &+\frac{256M\beta^2\lambda}{3b^3}+\frac{6144M^2\alpha\beta\lambda^2}{b^5}+\frac{1024M\beta^2\lambda^2}{3b^5},
   \end{split}
  \end{equation}
  for PPL and
  \begin{equation}
   \begin{split}
   \label{eq:M2}
   \hat{\alpha}\sim&\frac{4M}{b}-\frac{b\alpha^2}{2}\left(\frac{1}{u_{R}}+\frac{1}{u_{S}}\right)+bM\alpha^2-\frac{32M\lambda}{b^3}+\frac{48M^2\alpha\beta}{b}+\frac{8M\beta^2}{b}-\frac{768M^2\alpha\beta\lambda}{b^3}\\
   &-\frac{256M\beta^2\lambda}{3b^3}+\frac{43008M^2\alpha\beta\lambda^2}{b^5}+\frac{7168M\beta^2\lambda^2}{15b^5},
   \end{split}
  \end{equation}
  for PPR, respectively. Finally, we see that, when $\beta=0$, $\lambda=0$, and $\alpha^2=\frac{\Lambda}{3}$, the deflection angle will reduce to the form in  Kottler spacetime~\cite{ishihara2016gravitational}.
\section{Conclusions and discussions}
  In this paper, we have studied the equation of motion of the photon coupled to Weyl tensor by the geometric optics approximation and the corresponding shadow and weak gravitational lensing in a Brane-World black hole spacetime. Since the metric contains both dark energy parameter $\alpha$ and X-cold dark matter parameter $\beta$, we choose it as the background black hole spacetime. We find that the shadow and the gravitational lensing depend not only on the properties of background black hole spacetime, but also on the polarization of the coupled photon.
\par
  By calculating the shadow of the Brane-World black hole, we find that the photon sphere radius $r_{ph}$ increases with the coupling parameter $\lambda$ for PPL and decreases for PPR. And with the increases of $\beta$, $r_{ph}$ increases for two different coupled photons. Moreover, we have investigated the double shadow of the black hole as photons couple to the Weyl tensor, which does not appear in the non-coupling case where only a single shadow emerges. Combining with the supermassive central object in our Galaxy, we find the umbra of the black hole increases with the parameter $\beta$ and decreases with the coupling strength. The dependence of the penumbra on the parameter $\beta$ and the coupling strength is converse to that of the umbra.
\par
  We also study the gravitational deflection angle of light with finite-distance corrections in the weak deflection limit by means of a recent geometric method. Since the Brane-World black hole spacetime is nonasymptotically flat, the limit $bu_{R}\rightarrow 0$ and $bu_{S}\rightarrow 0$ is not allowed. However, it is not problematic because we can only observe a given star or a galaxy in finite distance from us.
\par
  Very recently, first images of the black hole M87 at the center of the Virgo A galaxy was obtained using the sub-millimeter ``EHT" based on the VLBI ~\cite{akiyama2019first}. We anticipate that future observations with highly improved techniques would be able to test our results by the observing the shadow and the deflection angle of the black hole.
\section*{Conflicts of Interest}
  The authors declare that there are no conflicts of interest regarding the publication of this paper.

\section*{Acknowledgments}
  We would like to thank the National Natural Science Foundation of China (Grant No.11571342) for supporting us on this work.

\section*{Appendix: Expressions of $\phi_{1}^{\ast}$, $\phi_{2}^{\ast}$, $\hat{\alpha}_{1}^{\ast}$ and $\hat{\alpha}_{2}^{\ast}$}
  By using Eq.~(\ref{eq:u}), $\phi_{RS}$  is expanded in a power series as Eq.~(\ref{eq:jifenjiaoPPL}) or Eq.~(\ref{eq:jifenjiaoPPM}), where
 \begin{align*}
  \phi_{1}^{\ast}=&\frac{bM\alpha^2}{2}\left[\frac{2-3b^2u_{R}^2}{\left(1-b^2u_{R}^2\right)^{\frac{3}{2}}}+\frac{2-3b^2u_{S}^2}{\left(1-b^2u_{S}^2\right)^{\frac{3}{2}}}\right]+b\alpha\beta\left(\frac{1}{\sqrt{1-b^2u_{R}^2}}+\frac{1}{\sqrt{1-b^2u_{S}^2}}\right)\\
  &+\frac{3M^2\alpha\beta}{b}\left[\frac{8-20b^2u_{R}^2+15b^4u_{R}^4}{\left(1-b^2u_{R}^2\right)^\frac{5}{2}}+\frac{8-20b^2u_{S}^2+15b^4u_{S}^4}{\left(1-b^2u_{S}^2\right)^\frac{5}{2}}\right]\\
  &+\frac{M\beta^2}{2b}\left[\frac{8-12b^2u_{R}^2}{\left(1-b^2u_{R}^2\right)^\frac{3}{2}}+\frac{8-12b^2u_{S}^2}{\left(1-b^2u_{S}^2\right)^\frac{3}{2}}\right]\\
  &+\frac{48M^2\alpha\beta\lambda}{b^3}\left[\frac{8-20b^2u_{R}^2+15b^4u_{R}^4}{\left(1-b^2u_{R}^2\right)^\frac{5}{2}}+\frac{8-20b^2u_{S}^2+15b^4u_{S}^4}{\left(1-b^2u_{S}^2\right)^\frac{5}{2}}\right]\\
  &+\frac{64M\beta^2\lambda}{3b^3}\left[\frac{2-3b^2u_{R}^2}{\left(1-b^2u_{R}^2\right)^\frac{3}{2}}+\frac{2-3b^2u_{S}^2}{\left(1-b^2u_{S}^2\right)^\frac{3}{2}}\right]\\
  &+\frac{384M^2\alpha\beta\lambda^2}{b^5}\left[\frac{8-20b^2u_{R}^2+15b^4u_{R}^4}{\left(1-b^2u_{R}^2\right)^\frac{5}{2}}+\frac{8-20b^2u_{S}^2+15b^4u_{S}^4}{\left(1-b^2u_{S}^2\right)^\frac{5}{2}}\right]\\
  &+\frac{256M\beta^2\lambda^2}{3b^5}\left[\frac{2-3b^2u_{R}^2}{\left(1-b^2u_{R}^2\right)^\frac{3}{2}}+\frac{2-3b^2u_{S}^2}{\left(1-b^2u_{S}^2\right)^\frac{3}{2}}\right],
 \end{align*}
 and
 \begin{align*}
  \phi_{2}^{\ast}=&\frac{bM\alpha^2}{2}\left[\frac{2-3b^2u_{R}^2}{\left(1-b^2u_{R}^2\right)^{\frac{3}{2}}}+\frac{2-3b^2u_{S}^2}{\left(1-b^2u_{S}^2\right)^{\frac{3}{2}}}\right]+b\alpha\beta\left(\frac{1}{\sqrt{1-b^2u_{R}^2}}+\frac{1}{\sqrt{1-b^2u_{S}^2}}\right)\\
  &+\frac{3M^2\alpha\beta}{b}\left[\frac{8-20b^2u_{R}^2+15b^4u_{R}^4}{\left(1-b^2u_{R}^2\right)^\frac{5}{2}}+\frac{8-20b^2u_{S}^2+15b^4u_{S}^4}{\left(1-b^2u_{S}^2\right)^\frac{5}{2}}\right]\\
  &+\frac{M\beta^2}{2b}\left[\frac{8-12b^2u_{R}^2}{\left(1-b^2u_{R}^2\right)^\frac{3}{2}}+\frac{8-12b^2u_{S}^2}{\left(1-b^2u_{S}^2\right)^\frac{3}{2}}\right]\\
  &-\frac{48M^2\alpha\beta\lambda}{b^3}\left[\frac{8-20b^2u_{R}^2+15b^4u_{R}^4}{\left(1-b^2u_{R}^2\right)^\frac{5}{2}}+\frac{8-20b^2u_{S}^2+15b^4u_{S}^4}{\left(1-b^2u_{S}^2\right)^\frac{5}{2}}\right]\\
  &-\frac{64M\beta^2\lambda}{3b^3}\left[\frac{2-3b^2u_{R}^2}{\left(1-b^2u_{R}^2\right)^\frac{3}{2}}+\frac{2-3b^2u_{S}^2}{\left(1-b^2u_{S}^2\right)^\frac{3}{2}}\right]\\
  &+\frac{2688M^2\alpha\beta\lambda^2}{b^5}\left[\frac{8-20b^2u_{R}^2+15b^4u_{R}^4}{\left(1-b^2u_{R}^2\right)^\frac{5}{2}}+\frac{8-20b^2u_{S}^2+15b^4u_{S}^4}{\left(1-b^2u_{S}^2\right)^\frac{5}{2}}\right]\\
  &+\frac{1792M\beta^2\lambda^2}{15b^5}\left[\frac{2-3b^2u_{R}^2}{\left(1-b^2u_{R}^2\right)^\frac{3}{2}}+\frac{2-3b^2u_{S}^2}{\left(1-b^2u_{S}^2\right)^\frac{3}{2}}\right].
 \end{align*}
 Similarly, after some calculations, we have
 \begin{align*}
  \hat{\alpha}_{1}^{\ast}=&\frac{bM\alpha^2}{2}\left[\frac{1-3b^2u_{R}^2}{\left(1-b^2u_{R}^2\right)^{\frac{3}{2}}}+\frac{1-3b^2u_{S}^2}{\left(1-b^2u_{S}^2\right)^{\frac{3}{2}}}\right]\\
  &+\frac{3M^2\alpha\beta}{b}\left[\frac{8-20b^2u_{R}^2+15b^4u_{R}^4}{\left(1-b^2u_{R}^2\right)^\frac{5}{2}}+\frac{8-20b^2u_{S}^2+15b^4u_{S}^4}{\left(1-b^2u_{S}^2\right)^\frac{5}{2}}\right]\\
  &+\frac{M\beta^2}{2b}\left[\frac{8-12b^2u_{R}^2}{\left(1-b^2u_{R}^2\right)^\frac{3}{2}}+\frac{8-12b^2u_{S}^2}{\left(1-b^2u_{S}^2\right)^\frac{3}{2}}\right]\\
  &+\frac{48M^2\alpha\beta\lambda}{5b^3}\left[\frac{8-20b^2u_{R}^2+15b^4u_{R}^4}{\left(1-b^2u_{R}^2\right)^\frac{5}{2}}+\frac{8-20b^2u_{S}^2+15b^4u_{S}^4}{\left(1-b^2u_{S}^2\right)^\frac{5}{2}}\right]\\
  &+\frac{64M\beta^2\lambda}{3b^3}\left[\frac{2-3b^2u_{R}^2}{\left(1-b^2u_{R}^2\right)^\frac{3}{2}}+\frac{2-3b^2u_{S}^2}{\left(1-b^2u_{S}^2\right)^\frac{3}{2}}\right]\\
  &+\frac{384M^2\alpha\beta\lambda^2}{b^5}\left[\frac{8-20b^2u_{R}^2+15b^4u_{R}^4}{\left(1-b^2u_{R}^2\right)^\frac{5}{2}}+\frac{8-20b^2u_{S}^2+15b^4u_{S}^4}{\left(1-b^2u_{S}^2\right)^\frac{5}{2}}\right]\\
  &+\frac{256M\beta^2\lambda^2}{b^5}\left[\frac{2-3b^2u_{R}^2}{\left(1-b^2u_{R}^2\right)^\frac{3}{2}}+\frac{2-3b^2u_{S}^2}{\left(1-b^2u_{S}^2\right)^\frac{3}{2}}\right],
 \end{align*}
 and
 \begin{align*}
  \hat{\alpha}_{2}^{\ast}=&\frac{bM\alpha^2}{2}\left[\frac{1-3b^2u_{R}^2}{\left(1-b^2u_{R}^2\right)^{\frac{3}{2}}}+\frac{1-3b^2u_{S}^2}{\left(1-b^2u_{S}^2\right)^{\frac{3}{2}}}\right]\\
  &+\frac{3M^2\alpha\beta}{b}\left[\frac{8-20b^2u_{R}^2+15b^4u_{R}^4}{\left(1-b^2u_{R}^2\right)^\frac{5}{2}}+\frac{8-20b^2u_{S}^2+15b^4u_{S}^4}{\left(1-b^2u_{S}^2\right)^\frac{5}{2}}\right]\\
  &+\frac{M\beta^2}{2b}\left[\frac{8-12b^2u_{R}^2}{\left(1-b^2u_{R}^2\right)^\frac{3}{2}}+\frac{8-12b^2u_{S}^2}{\left(1-b^2u_{S}^2\right)^\frac{3}{2}}\right]\\
  &-\frac{48M^2\alpha\beta\lambda}{5b^3}\left[\frac{8-20b^2u_{R}^2+15b^4u_{R}^4}{\left(1-b^2u_{R}^2\right)^\frac{5}{2}}+\frac{8-20b^2u_{S}^2+15b^4u_{S}^4}{\left(1-b^2u_{S}^2\right)^\frac{5}{2}}\right]\\
  &-\frac{64M\beta^2\lambda}{3b^3}\left[\frac{2-3b^2u_{R}^2}{\left(1-b^2u_{R}^2\right)^\frac{3}{2}}+\frac{2-3b^2u_{S}^2}{\left(1-b^2u_{S}^2\right)^\frac{3}{2}}\right]\\
  &+\frac{2688M^2\alpha\beta\lambda^2}{b^5}\left[\frac{8-20b^2u_{R}^2+15b^4u_{R}^4}{\left(1-b^2u_{R}^2\right)^\frac{5}{2}}+\frac{8-20b^2u_{S}^2+15b^4u_{S}^4}{\left(1-b^2u_{S}^2\right)^\frac{5}{2}}\right]\\
  &+\frac{1792M\beta^2\lambda^2}{5b^5}\left[\frac{2-3b^2u_{R}^2}{\left(1-b^2u_{R}^2\right)^\frac{3}{2}}+\frac{2-3b^2u_{S}^2}{\left(1-b^2u_{S}^2\right)^\frac{3}{2}}\right].
 \end{align*}

\section*{References}

 \bibliographystyle{unsrt}
 \bibliography{ref}
\end{document}